\documentclass[aps,prc,twocolumn,superscriptaddress,floatfix,amsmath,amssymb,nofootinbib,longbibliography]{revtex4-2}

\usepackage[utf8]{inputenc}
\usepackage[T1]{fontenc}

\usepackage{siunitx}
\usepackage{xcolor}
\usepackage{graphicx}

\usepackage[colorlinks=true,citecolor=blue,linkcolor=blue,urlcolor=blue]{hyperref}

\usepackage{scalefnt}
\usepackage{times,txfonts}

\usepackage{nicefrac}
\usepackage{mathtools}
\usepackage{tabularx}
\usepackage{booktabs}
\usepackage{float}
\newcolumntype{L}[1]{>{\raggedright\let\newline\\\arraybackslash\hspace{0pt}}m{#1}}

\usepackage{braket}

\newcommand{\vnn}[4]{\operatorname{}\bra{#1#2}V_{\text{NN}}\ket{#3#4}}
\newcommand{\vtn}[6]{\operatorname{}\bra{{#1}{#2}{#3}}V_{\text{3N}}\ket{#4#5#6}}
\newcommand{\be}{\begin{equation}}
\newcommand{\ee}{\end{equation}}

\newcommand{\fmi}{\, \text{fm}^{-1}}
\newcommand{\fmiq}{\, \text{fm}^{-3}}

\newcommand{\mev}{\, \text{MeV}}

\begin{document}

\title{Equation of state and Fermi liquid properties of dense matter\\ based on chiral effective field theory interactions}

\author{F. Alp}
\email{faruk_musa.alp@tu-darmstadt.de}
\affiliation{Technische Universit\"at Darmstadt, Department of Physics, D-64289 Darmstadt, Germany} 
\affiliation{ExtreMe Matter Institute EMMI, GSI Helmholtzzentrum f\"ur Schwerionenforschung GmbH, D-64291 Darmstadt, Germany}
\affiliation{Max-Planck-Institut f\"ur Kernphysik, Saupfercheckweg 1, D-69117 Heidelberg, Germany}

\author{Y. Dietz}
\email{yannick.dietz@tu-darmstadt.de}
\affiliation{Technische Universit\"at Darmstadt, Department of Physics, D-64289 Darmstadt, Germany} 
\affiliation{ExtreMe Matter Institute EMMI, GSI Helmholtzzentrum f\"ur Schwerionenforschung GmbH, D-64291 Darmstadt, Germany}

\author{K. Hebeler}
\email{kai.hebeler@physik.tu-darmstadt.de}
\affiliation{Technische Universit\"at Darmstadt, Department of Physics, D-64289 Darmstadt, Germany} 
\affiliation{ExtreMe Matter Institute EMMI, GSI Helmholtzzentrum f\"ur Schwerionenforschung GmbH, D-64291 Darmstadt, Germany}
\affiliation{Max-Planck-Institut f\"ur Kernphysik, Saupfercheckweg 1, D-69117 Heidelberg, Germany}

\author{A. Schwenk}
\email{schwenk@physik.tu-darmstadt.de}
\affiliation{Technische Universit\"at Darmstadt, Department of Physics, D-64289 Darmstadt, Germany} 
\affiliation{ExtreMe Matter Institute EMMI, GSI Helmholtzzentrum f\"ur Schwerionenforschung GmbH, D-64291 Darmstadt, Germany}
\affiliation{Max-Planck-Institut f\"ur Kernphysik, Saupfercheckweg 1, D-69117 Heidelberg, Germany}

\begin{abstract}
We present results for the equation of state of symmetric nuclear matter and pure neutron matter obtained in many-body-perturbation theory (MBPT) up to third order, based on various chiral two- and three-nucleon interactions used in ab initio calculations of nuclei. We extract equation of state properties, such as the incompressibility and the symmetry energy, and discuss estimates of the theoretical uncertainties due to neglected higher-order contributions in the MBPT expansion as well as the chiral effective field theory expansion. 
In addition, we discuss the Fermi liquid approach to nuclear matter. We calculate all two- and three-nucleon contributions to the quasiparticle interaction up to second order in MBPT and present results for the Landau parameters, effective mass, and speed of sound for pure neutron matter.
\end{abstract}

\maketitle

\section{Introduction}

Ab initio calculations of the nuclear equation of state (EOS) have seen tremendous progress over the last decade. This is based on the development of ab initio many-body frameworks~\cite{HebelerSchwenk2010,Tews13N3LO,Holt13PPNP,Carb13nm,Hage14ccnm,Coraggio_et_al_2014,PhysRevC.89.064009,PhysRevC.92.015801,Lynn16QMC3N,Dris16asym,Ekst17deltasat,Drischler2019,CarboneSchwenk2019,Lu2020,Keller2021,Keller2023,Marino:2024tfp} and the derivation of new nuclear interactions from chiral effective field theory (EFT)~\cite{Epelbaum_et_al_2009,Machleidt_Entem_2011,Hammer2020}. This makes it possible to access various EOS quantities and quantify their theoretical uncertainties in a systematic way~\cite{Hebeler2015,Lynn2019,Drischler2021ARNPS}. Ab initio calculations can provide constraints for various observables such as, the pressure of neutron star matter, the speed of sound, the effective mass, and also transport properties, which represent key inputs for astrophysical studies of core-collapse supernovae, neutron star structure and mergers (see, e.g., Refs.~\cite{Yasin:2018ckc,Schneider:2019shi,Lattimer2021,Jacobi:2023olu}). Many of these EOS quantities are also accessible in Fermi liquid theory~\cite{Landau:3,Landau:5,Landau:8,BaymPethick:1991,Friman:2012ft}. In this framework, observables can be directly related to the Landau parameters, which parametrize the interaction between quasiparticles, the fundamental degrees of freedom in Fermi liquid theory. In Refs.~\cite{Schwenk:2002fq,Schwenk:2003bc,Holt:2011,Holt:2013} first calculations of Landau parameters were presented based on modern nucleon-nucleon (NN) and three-nucleon (3N) interactions. In this work, we will explore Fermi liquid properties from many-body perturbation theory (MBPT) including all contributions from NN and 3N interactions up to second order.
\nopagebreak
In recent years, several new nuclear interactions have been developed. While previous chiral interactions have been quite successful at correctly describing the ground-state energies of light and medium-mass nuclei, experimental charge and mass radii have been more difficult to reproduce theoretically~\cite{Simonis:2017dny}. Recently, new interactions have been introduced within $\Delta$-full chiral EFT~\cite{Ekst17deltasat,Jian20N2LOGO}. Additionally, new low-resolution interactions have been developed by~\citet{Arthuis:2024mnl} for calculations of medium-mass and heavy nuclei using an adapted fitting strategy of the 3N low-energy constants for an improved description of radii compared to previously derived low-resolution interactions~\cite{Hebeler_et_al2011}, including the 1.8/2.0~( Entem and Machleidt, or EM) interaction that has been widely used in ab initio calculations of nuclei up to $^{208}$Pb (see, e.g., Refs.~\cite{Stroberg:2019bch,Hebeler2021,Miyagi2022,Hebeler:2022aui,Tichai:2023epe}).
\newline\indent
The goal of this work is to provide EOS results for these new interactions and to compare EOS calculations for interactions derived in $\Delta$-less chiral EFT with those derived from $\Delta$-full chiral EFT. For this we compute the energy per particle of pure neutron matter (PNM) and symmetric nuclear matter (SNM) for a set of new and established $\Delta$-less interactions as well as the mentioned $\Delta$-full interactions. We employ a Gaussian process to obtain statistical uncertainties associated with the truncation of the chiral EFT expansion within a Bayesian approach~\cite{Mele17bayerror,Melendez2019,Drischler2020PRL,Drischler2020} and discuss the sensitivity of various EOS observables to the employed interactions.
\newline\indent
In the second part of this paper, we calculate the quasiparticle interaction of Fermi liquid theory by including all contributions from NN and 3N interactions up to second order in MBPT. We present results for the Landau parameters and resulting quantities, such as the effective mass and the speed of sound for PNM, and discuss results for different interactions as well as uncertainties from the MBPT and chiral expansion.
\newline\indent
This paper is organized as follows: In Sec.~\ref{chap:EOS} we discuss our calculations of the EOS of PNM and SNM. We start by introducing MBPT as our many-body method, then we specify the employed interactions and discuss our results and uncertainties. In Sec.~\ref{chap:FLT} we review the framework for the calculation of Landau parameters in MBPT and discuss in detail the contributing diagrams, the relevant relations for the studied observables, and results for the different interactions. While our MBPT framework has been developed up to third order including a full incorporation of 3N contributions, our Fermi liquid calculations include contributions only up to second order due to the significant increase in the number of diagrams.\ Finally, we summarize our findings and give an outlook in Sec.~\ref{chap:summary}.

\section{Equation of state}
\label{chap:EOS}

\begin{table*}[t]
\begin{tabular}{l|S[table-number-alignment = center]S[table-number-alignment = center]S[table-number-alignment = center]S[table-number-alignment = center]S[table-number-alignment = center]|cc}
Interaction & \text{$c_1$} & \text{$c_3$} & \text{$c_4$} & \text{$c_D$} & \text{$c_E$} & \hspace*{0.5mm} $\Lambda_{\text{3N}}$ \hspace*{0.5mm} & $n_{\text{exp}}$ \\
\hline
N$^2$LO EMN 450 & -0.74 & -3.61 & 2.44 & 2.5 & 0.103 & 450 & $4$ \\
N$^2$LO EMN 500 & -0.74 &-3.61 & 2.44 &-1.50 &-0.61 & 500 & 4 \\
1.8/2.0 (EM) & -0.81 & -3.2 & 5.4 & 1.264 & -0.12 & 2.0 & 4 \\
1.8/2.0 (EM7.5) & -0.81 & -3.2 & 5.4 & 7.5 & 0.942 & 2.0 & 4 \\
1.8/2.0 (sim7.5) & 0.27 & -3.56 & 3.644 & 7.5 & 1.081 & 2.0 & 3 \\$\Delta$NLO$_{\text{GO}}\,(450)$ & 0 & -2.972 & 1.486 & 0 & 0 & 450 & 3 \\
$\Delta$NNLO$_{\text{GO}}\,(450)$\, & -0.74 & -3.622 & 2.446 & -0.454 & -0.186 & 450 & 3 \\
$\Delta$NNLO$_{\text{GO}}\,(394)$\, & -0.74 & -3.622 & 2.446 & 0.081 & -0.002 & 394 & 4
\end{tabular}
\caption{Three-nucleon low-energy-coupling (LEC) values and regulator parameters for all interactions used in this work. The LECs $c_1$, $c_3$, and $c_4$ are given in units of $\text{GeV}^{-1}$. The 3N cutoff $\Lambda_{\text{3N}}$ is given in MeV, except for the 1.8/2.0 interactions where it is given in $\text{fm}^{-1}$. For the $\Delta$-full interactions, the $c_i$ values include also the contributions from the $\Delta$.}
\label{tab:LECs}
\end{table*}
\subsection{Many-body framework}
\label{chap:EOS_MBPT}

For our EOS calculations of nuclear matter we employ MBPT. In previous works, we have developed a versatile framework based on Monte Carlo (MC) sampling techniques that evaluates individual MBPT diagrams involving NN and 3N interactions in an efficient way~\cite{Drischler2019,Keller2021,Keller2023}. This framework allows one to compute the EOS for arbitrary temperatures and isospin asymmetries based on general NN and 3N interactions up to third order in MBPT. In this work, we limit ourselves to results for pure neutron matter (PNM) and symmetric nuclear matter (SNM) at zero temperature. In practice these calculations are performed in the grand-canonical ensemble at finite temperature by choosing a very small temperature of $T=10^{-3}$\,MeV. We have checked that this choice leads to results in essentially perfect agreement with results of the framework at zero temperature in the canonical ensemble.

We follow \citet{Keller2021,Keller2023} and consider the perturbative expansion of the grand-canonical potential around a Hartree-Fock (HF) reference state. This choice leads to a natural MBPT convergence pattern for all studied interactions and, importantly, cancels the anomalous diagrams which arise starting at second order in the perturbation series at finite temperature. Individual diagrams in the MBPT expansion are evaluated using a MC integration algorithm~\cite{Hahn05CUBA,Hahn16CUBApara} to a precision of 5\,keV for the energy per particle. In our previous work, we have shown that this amounts to a numerical uncertainty, which is significantly smaller than the chiral uncertainties and MBPT uncertainties~\cite{Keller2023}. Our calculations make use of partial-wave decomposed NN potentials where we include all partial waves up to total angular momentum $J \leq 12$ for all interactions studied. This makes it possible to use general NN interactions, including similarity renormalization group (SRG) evolved interactions, such as the low-resolution interactions 1.8/2.0 (see next section). We have checked that the NN HF energies are converged to less than 10\,keV at nuclear saturation density $n_0 = 0.16 \fmiq$ and less than 30\,keV at $2 n_0$. This is on the scale of less than a per mille with respect to the total interaction contribution.

For this work, we include 3N interactions in a single-particle operator basis as in Ref.~\cite{Drischler2019}, i.e., we evaluate the $V_\text{3N}$ interaction matrix elements that arise in the MBPT expansion for the 3N operators directly in single-particle states $\ket{{\bf k}_i, \boldsymbol{\sigma}_i, \boldsymbol{\tau}_i}$, rather than going through a 3N partial-wave expansion as in Ref.~\cite{Keller2021,Keller2023}. This allows for a fast and efficient evaluation of the 3N contributions and reduces the runtime by a factor of about 50 for most of the MBPT diagrams compared to using a partial-wave-decomposed form for 3N interactions. These efficiency gains are possible for most 3N interactions, but not for all,
e.g., not for SRG-evolved 3N interactions that are available only in partial-wave basis. In addition, for higher-order 3N interactions or 3N interactions with more complicated regularization schemes, the use of 3N partial wave matrix elements is helpful, as the 3N partial-wave matrix elements are precomputed.\ The given boost in efficiency combined with the MC evaluation technique permits us to evaluate individual diagrams up to third order in a matter of minutes, with a full EOS calculation requiring approximately 3000 CPU hours. The efficient evaluation is especially important for higher orders in MBPT, where both the number of diagrams as well as the dimension of the phase space integrals in each diagram increases rapidly.

We have benchmarked our implementation against previous calculations that used partial-wave-decomposed 3N interactions~\cite{Keller2021,Keller2023} as well as studies using the single-particle basis~\cite{Drischler2019}. We furthermore validated our MBPT implementation by benchmarking against MBPT calculations in a finite box using a discrete basis representation~\cite{Marino:2024tfp}.

\subsection{Hamiltonians}
\label{chap:EOS_Hamiltonians}

In this work, we employ three different families of chiral EFT interactions, each including contributions from NN and 3N interactions. The first set consists of the next-to-next-to-leading order (N$^2$LO) NN interactions of Entem, Machleidt, and Nosyk (EMN)~\cite{EMN2017} with a cutoff $\Lambda = 450$ and $500 \mev$. These are labeled N$^2$LO EMN 450 and N$^2$LO EMN 500. The low-energy couplings (LECs) $c_D$ and $c_E$ of the corresponding 3N interactions at N$^2$LO are determined by fits to the $^3$H binding energy and the saturation point of symmetric nuclear matter~\cite{Drischler2019}. All 3N interactions used in this work employ a nonlocal regulator function of the form $f_{\Lambda}(p,q) = \exp\left[-((p^2 + 3/4 q^2)/\Lambda_{\text{3N}}^2)^{n_{\text{exp}}} \right]$, where $p$ and $q$ denote the Jacobi momenta. The specific LEC values and the cutoff exponent $n_{\text{exp}}$ for all interactions used in this work are summarized in Table~\ref{tab:LECs}. 

The second set consists of the $\Delta$-full interactions at next-to-leading order (NLO) and N$^2$LO for cutoffs $\Lambda = 394$ and $450 \mev$~\cite{Ekst17deltasat,Jian20N2LOGO}. These are labeled $\Delta$NLO$_{\text{GO}}\,(450)$,
$\Delta$NNLO$_{\text{GO}}\,(450)$, and
$\Delta$NNLO$_{\text{GO}}\,(394)$. Especially the latter Hamiltonian has been used in calculations up to heavy nuclei; see, e.g., Refs.~\cite{Hu2022,Arthuis:2024mnl}. Moreover, these have been recently employed in nuclear matter calculations at zero temperature using coupled-cluster, MBPT and self-consistent Green's function methods~\cite{Jian20N2LOGO,Marino:2024tfp}. 

Finally, we include the low-resolution interactions 1.8/2.0 (EM) of Ref.~\cite{Hebeler_et_al2011} as well as 1.8/2.0 (EM7.5) and 1.8/2.0 (sim7.5) of Ref.~\cite{Arthuis:2024mnl}.
These interactions are derived by evolving the N$^3$LO NN potential of Entem and Machleidt (EM)~\cite{Ente03EMN3LO} with a cutoff $\Lambda = 500 \mev$ or the N$^2$LO$_{\text{sim}}$ NN interaction (sim) \cite{Carl15sim} with $\Lambda = 550 \mev$ to the resolution scale $\lambda_{\text{SRG}} = 1.8 \fmi$ using the SRG. The 3N LECs $c_D$ and $c_E$ are determined by fits to few- and many-body observables at a cutoff scale of $\Lambda_{\text{3N}} = 2.0 \fmi$. For 1.8/2.0 (EM) the $^3$H binding energy and $^4$He  radius were used~\cite{Hebeler_et_al2011}, while for 1.8/2.0 (EM7.5) and 1.8/2.0 (sim7.5) the 3N interaction was fitted to the $^3$H binding energy as well as the ground-state energy and charge radius of $^{16}$O~\cite{Arthuis:2024mnl}. In a recent study, the new 1.8/2.0 (EM7.5) and 1.8/2.0 (sim7.5) interactions were successfully applied to describe bulk properties and neutron skins of nuclei across the nuclear chart~\cite{Arthuis:2024mnl}.

\begin{figure}[t!]
    \includegraphics[width=\columnwidth]{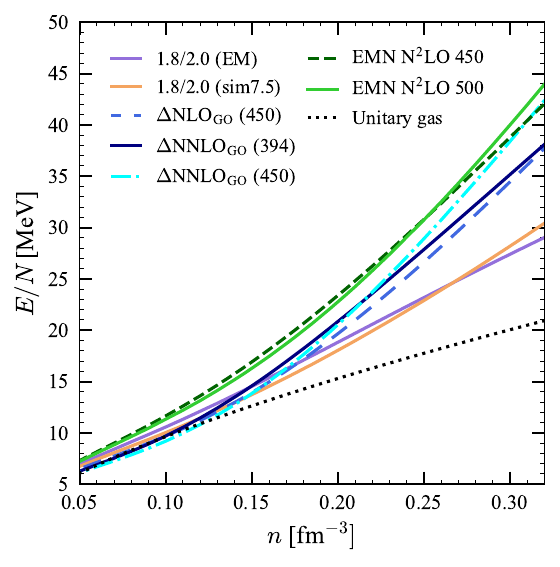}  
    \caption{Energy per particle $E/N$ of PNM as a function of particle density $n$ based on different NN and 3N interactions (see main text). Note that the results for the interactions 1.8/2.0 (EM) and 1.8/2.0 (EM7.5) are identical since the short-range terms $c_D$ and $c_E$ do not contribute to neutron matter for nonlocal regulators. The dotted line shows the energy per particle of the unitary Fermi gas for comparison~\cite{Ku2012}.}
    \label{fig:EOS_PNM}
\end{figure}

\begin{figure}[t!]
    \includegraphics[width=\columnwidth]{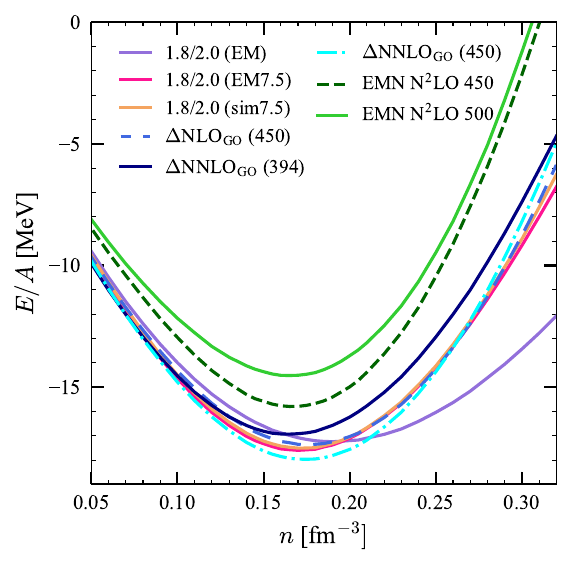}
    \caption{Energy per particle $E/A$ of SNM as a function of particle density $n$ based on the same NN and 3N interactions as in Fig.~\ref{fig:EOS_PNM}.}
    \label{fig:EOS_SNM}
\end{figure}

\subsection{Results}
\label{chap:EOS_Results}

\subsubsection{Neutron matter and symmetric matter}
Figure~\ref{fig:EOS_PNM} shows our results for the energy per particle $E/N$ of neutron matter at third order in MBPT. As in Ref.~\cite{Keller2023}, all our calculations are based on an expansion around a HF reference state (using a HF spectrum for diagrams beyond HF) and include 3N contributions up to second order fully, while the third-order diagrams are only included with 3N contributions as normal-ordered two-body terms, i.e., neglecting residual 3N operators at third order. We find that the results for all studied interactions agree within about $3 \mev$ at $n_0$. At higher densities the results for the low-resolution interactions 1.8/2.0 are lower than for all the other interactions. The SRG-evolved interactions 1.8/2.0 (EM) and 1.8/2.0 (EM7.5) result in the same energy per particle for PNM, as the short-range terms $c_D$ and $c_E$ in PNM vanish for nonlocal regulators~\cite{HebelerSchwenk2010}. We note that the 1.8/2.0 (sim7.5) interaction produces results that are very close to 1.8/2.0 (EM), which has also been observed in finite nuclei~\cite{Arthuis:2024mnl}.

We additionally show the energy of the unitary Fermi gas for comparison, which may serve as a lower bound for the energy of PNM~\cite{Tews_2017}. We note that the N$^2$LO $\Delta$-full interactions considered here violate this lower bound for densities below $n_0$. This is likely caused by too attractive spin-triplet $P$-wave contributions. With increasing density, the interactions become significantly more repulsive around saturation density due to 3N contributions. Finally, we note that the EMN N$^2$LO 450 and EMN N$^2$LO 500 interactions give the largest energies, while both interactions provide very similar results.

\begin{figure*}[t]
    \centering
    \includegraphics[width = 0.85\textwidth]{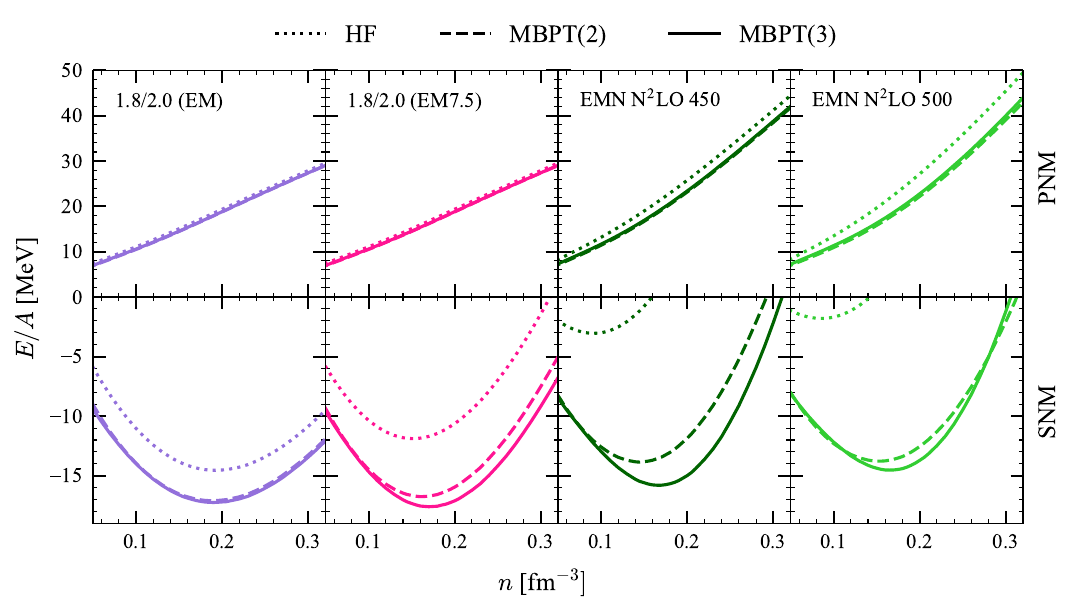}
    \includegraphics[width = 0.85\textwidth]{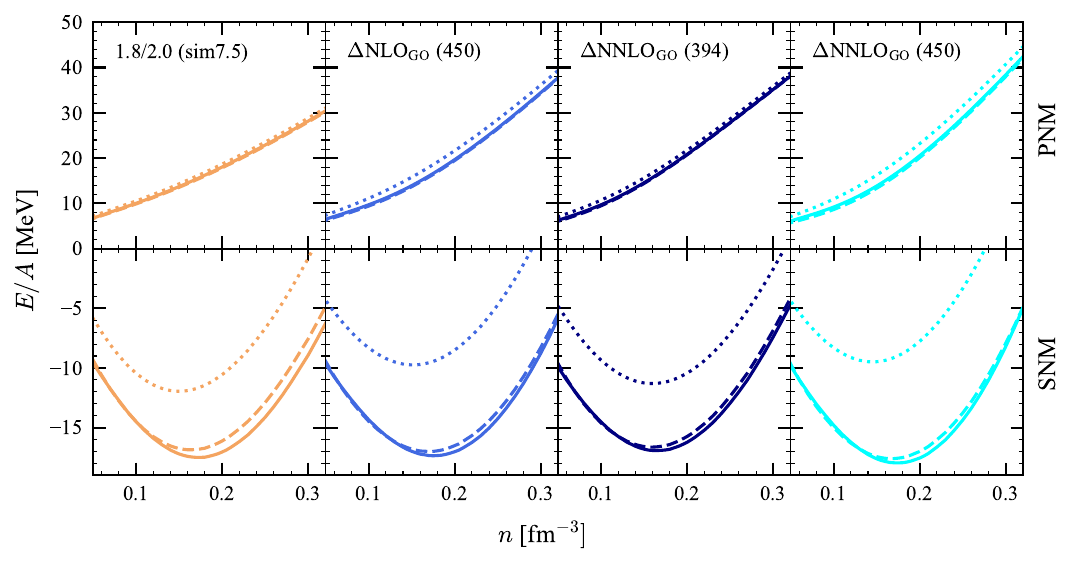}
    \caption{Energy per particle of neutron matter (upper rows in both panels) and symmetric matter (lower rows) for different NN and 3N interactions at different orders in the MBPT expansion. Results are shown at the HF level (dotted lines), including MBPT(2) (dashed lines) and MBPT(3) (solid lines).}
    \label{fig:EOS_MBPT}
\end{figure*}

Figure \ref{fig:EOS_SNM} shows the corresponding results for SNM. The results based on interactions within one family are found to share certain trends. For instance, both the N$^2$LO EMN 450 and N$^2$LO EMN 500 show a higher energy per particle over the whole density range than the $\Delta$-full interactions and the SRG-evolved interactions 1.8/2.0. The higher saturation energy for the N$^2$LO EMN 450 and N$^2$LO EMN 500 is by construction (see Ref.~\cite{Drischler2019}), while all other interactions except for 1.8/2.0 (EM) have been optimized to describe also nuclei, leading to a more realistic saturation point. While the 1.8/2.0 (EM) interaction yields a good saturation energy, it saturates at too high density, about $0.02 \fmiq$ higher than the 1.8/2.0 (EM7.5) interaction (see also Ref.~\cite{Simonis:2017dny}). For the new 1.8/2.0 (EM7.5) and 1.8/2.0 (sim7.5) interactions, the improved saturation properties are caused by the larger $c_D$ term necessary to reproduce radii  of nuclei correctly~\cite{Arthuis:2024mnl}. For higher density, we find that the new 1.8/2.0 (EM7.5) and 1.8/2.0 (sim7.5) interactions are similar to the $\Delta$-full interactions, while the 1.8/2.0 (EM) interaction leads to a softer EOS in SNM.

\subsubsection{Many-body and chiral EFT uncertainties}

For an estimate of the theoretical uncertainties of our results we need to take into account various contributions:
\begin{enumerate}
    \item Truncation effects due to using a finite number of partial-wave channels for the NN interactions.
    \item Finite accuracy of the MC sampler when evaluating individual MBPT diagrams.
    \item Uncertainties due to neglected higher-order diagrams in the MBPT expansion.
    \item Uncertainties due to neglected higher-order contributions in the chiral EFT expansion of NN, 3N and higher-body interactions.
\end{enumerate}
In Sec.~\ref{chap:EOS_MBPT} we already discussed the numerical uncertainties from the MC solver as well as the uncertainties due to the expansion of the NN interaction into partial waves. Both of these contributions can be considered negligible compared to the other two contributions.

\begin{table*}[t]
\sisetup{table-format=3.0, table-number-alignment=center, table-column-width=18mm}
\begin{tabular}
{l|c|SSSSSSS}
\text{Interaction} & \hspace*{2.5mm} $n$ \hspace*{2.5mm} & \text{$E^{(1)}_{\text{NN}}$} & \text{$E^{(1)}_{\text{3N}}$} & \text{$E^{(2)}_{\text{NN}}$} & \text{$E^{(2)}_{\text{NN+3N}}$} & \text{$E^{(2)}_{\text{3N-res}}$} & \text{$E^{(3)}_{\text{NN}}$} & \text{$E^{(3)}_{\text{NN+3N}}$} \\
\hline
N$^2$LO EMN 450 & $n_0$ & -21.12 & 6.29 & -1.66 & -0.68 & -0.01 & 0.24 & -0.01 \\ 
& $2 n_0$ & -33.40 & 22.20 & -1.94 & -0.76 & -0.00 & 0.19 & 0.13 \\
\hline
N$^2$LO EMN 500 & $n_0$ & -20.51 & 6.49 & -2.59 & -1.48 & -0.05 & 0.42 & 0.09 \\ 
& $2 n_0$ & -31.64 & 25.50 & -3.81 & -2.74 & -0.02 & 0.42 & 0.61 \\
\hline
1.8/2.0 (EM) & $n_0$ & -24.36 & 5.20 & -0.49 & -0.12 & -0.00 & 0.08 & -0.02 \\ 
& $2 n_0$ & -40.98 & 14.86 & -0.48 & -0.06 & -0.00 & 0.06 & -0.02 \\
\hline
1.8/2.0 (sim7.5) & $n_0$ & -24.46 & 4.55 & -0.67 & -0.12 & -0.00 & 0.17 & -0.00 \\ 
& $2 n_0$ & -37.52 & 12.95 & -0.78 & -0.06 & -0.00 & 0.20 & -0.00 \\
\hline
$\Delta$NLO$_{\textrm{GO}}\,(450)$ & $n_0$ & -22.67 & 4.47 & -1.53 & -0.60 & -0.01 & 0.24 & -0.00 \\ 
& $2 n_0$ & -31.58 & 15.28 & -1.14 & -0.60 & -0.00 & 0.11 & 0.04 \\
\hline
$\Delta$NNLO$_{\textrm{GO}}\,(450)$\, & $n_0$ & -23.49 & 5.94 & -2.15 & -0.81 & -0.01 & 0.46 & 0.03 \\ 
& $2 n_0$ & -30.99 & 19.97 & -1.77 & -0.82 & -0.00 & 0.26 & 0.12 \\
\hline
$\Delta$NNLO$_{\textrm{GO}}\,(394)$\, & $n_0$ & -24.05 & 5.77 & -0.99 & -0.19 & -0.00 & 0.17 & -0.02 \\
& $2 n_0$ & -33.36 & 16.53 & -0.67 & -0.08 & -0.00 & 0.06 & -0.01 \\
\end{tabular}
\caption{MBPT contributions to the energy per particle (in MeV) of PNM at $n_0 = 0.16 \fmiq$ and $2 n_0 = 0.32 \fmiq$ for the employed interactions. The columns give the NN and 3N HF energies ($E^{(1)}_{\text{NN}}$ and $E^{(1)}_{\text{3N}}$), the MBPT(2) corrections with NN interactions only ($E^{(2)}_{\text{NN}}$), involving 3N contributions at the normal-ordered two-body level ($E^{(2)}_{\text{NN+3N}}$), and the residual 3N-3N diagram ($E^{(2)}_{\text{3N-res}}$), and MBPT(3) corrections with NN only ($E^{(3)}_{\text{NN}}$) and including normal-ordered two-body contributions from 3N ($E^{(3)}_{\text{NN+3N}}$). Note that all energies are energies per particle. Numbers are rounded to the $10\,\text{keV}$ level, given the accuracy of our MC solver.
\label{tab:PNM_MBPTdata}}
\end{table*}

\begin{table*}[t]
\sisetup{table-format=3.0, table-number-alignment=center, table-column-width=18mm}
\begin{tabular}
{l|c|SSSSSSS}
\text{Interaction} & \hspace*{2.5mm} $n$ \hspace*{2.5mm} & \text{$E^{(1)}_{\text{NN}}$} & \text{$E^{(1)}_{\text{3N}}$} & \text{$E^{(2)}_{\text{NN}}$} & \text{$E^{(2)}_{\text{NN+3N}}$} & \text{$E^{(2)}_{\text{3N-res}}$} & \text{$E^{(3)}_{\text{NN}}$} & \text{$E^{(3)}_{\text{NN+3N}}$} \\
\hline
\text{N$^2$LO EMN 450} & \text{$n_0$} & -34.14 & 12.10 & -6.22 & -7.12 & -0.42 & 0.23 & -2.31 \\ 
& \text{$2 n_0$} & -56.28 & 51.77 & -6.14 & -18.72 & -0.35 & 0.33 & -3.60 \\
\hline
N$^2$LO EMN 500 & $n_0$ & -31.61 & 10.83 & -8.59 & -5.82 & -0.66 & 0.27 & -1.04 \\ 
& $2 n_0$ & -51.66 & 46.16 & -9.13 & -18.40 & -0.80 & 0.32 & 1.64 \\
\hline
1.8/2.0 (EM) & $n_0$ & -43.85 & 7.61 & -2.31 & -0.24 & -0.10 & 0.04 & -0.14 \\ 
& $2 n_0$ & -73.61 & 28.99 & -1.88 & -0.49 & -0.04 & 0.16 & -0.32 \\
\hline
1.8/2.0 (EM7.5) & $n_0$ & -43.85 & 9.90 & -2.25 & -2.53 & -0.12 & 0.04 & -0.82 \\ 
& $2 n_0$ & -73.61 & 39.71 & -2.01 & -4.18 & -0.06 & 0.18 & -1.93 \\
\hline
1.8/2.0 (sim7.5) & $n_0$ & -43.65 & 9.65 & -2.24 & -2.59 & -0.14 & 0.10 & -0.66 \\ 
& $2 n_0$ & -72.24 & 39.15 & -1.96 & -4.88 & -0.07 & 0.31 & -1.70 \\
\hline
$\Delta$NLO$_{\textrm{GO}}\,(450)$ & $n_0$ & -38.87 & 7.05 & -4.37 & -2.77 & -0.13 & 0.27 & -0.54 \\ 
& $2 n_0$ & -60.89 & 30.28 & -3.76 & -6.10 & -0.11 & 0.17 & -0.56 \\
\hline
$\Delta$NNLO$_{\textrm{GO}}\,(450)$\, & $n_0$ & -41.12 & 9.64 & -4.91 & -3.08 & -0.22 & 0.35 & -0.62 \\ 
& $2 n_0$ & -67.49 & 39.44 & -4.37 & -7.37 & -0.19 & 0.25 & -0.33 \\
\hline
$\Delta$NNLO$_{\textrm{GO}}\,(394)$\, & $n_0$ & -42.80 & 9.38 & -3.71 & -1.56 & -0.07 & 0.13 & -0.42 \\ 
& $2 n_0$ & -71.61 & 37.49 & -2.68 & -2.41 & -0.03 & 0.11 & -0.66 \\
\end{tabular}
\caption{Same as Table~\ref{tab:PNM_MBPTdata}, but for SNM. Note that for PNM the 1.8/2.0 (EM7.5) interaction yields the same results as the 1.8/2.0 (EM) interaction, as the 3N terms $c_D$ and $c_E$ do not contribute to PNM.
\label{tab:SNM_MBPTdata}}
\end{table*}

Figure~\ref{fig:EOS_MBPT} shows the MBPT convergence behavior for the different NN and 3N interactions considered in this work. For PNM (upper rows in both panels) we observe a good convergence for all Hamiltonians studied. Unsurprisingly, the low-resolution interactions 1.8/2.0 exhibit the fastest MBPT convergence, while the EMN N$^2$LO 500 interaction shows the slowest perturbative convergence. At saturation density $n_0$, we find a MBPT(2) correction of about $E^{(2)} = -0.6 \mev$ for 1.8/2.0 (EM), whereas the N$^2$LO EMN 500 gives $E^{(2)} = -4.1 \mev$. At third order, the contribution for the latter amounts to $E^{(3)} = 0.5 \mev$, whereas the MBPT(3) correction for the 1.8/2.0 (EM) interaction is only $E^{(3)} = -0.06 \mev$. The individual contributions at different orders in the MBPT expansion for all studied Hamiltonians are summarized in Tables~\ref{tab:PNM_MBPTdata} and  \ref{tab:SNM_MBPTdata}.

\begin{figure*}[t]
    \centering
    \includegraphics[width=0.8\textwidth]{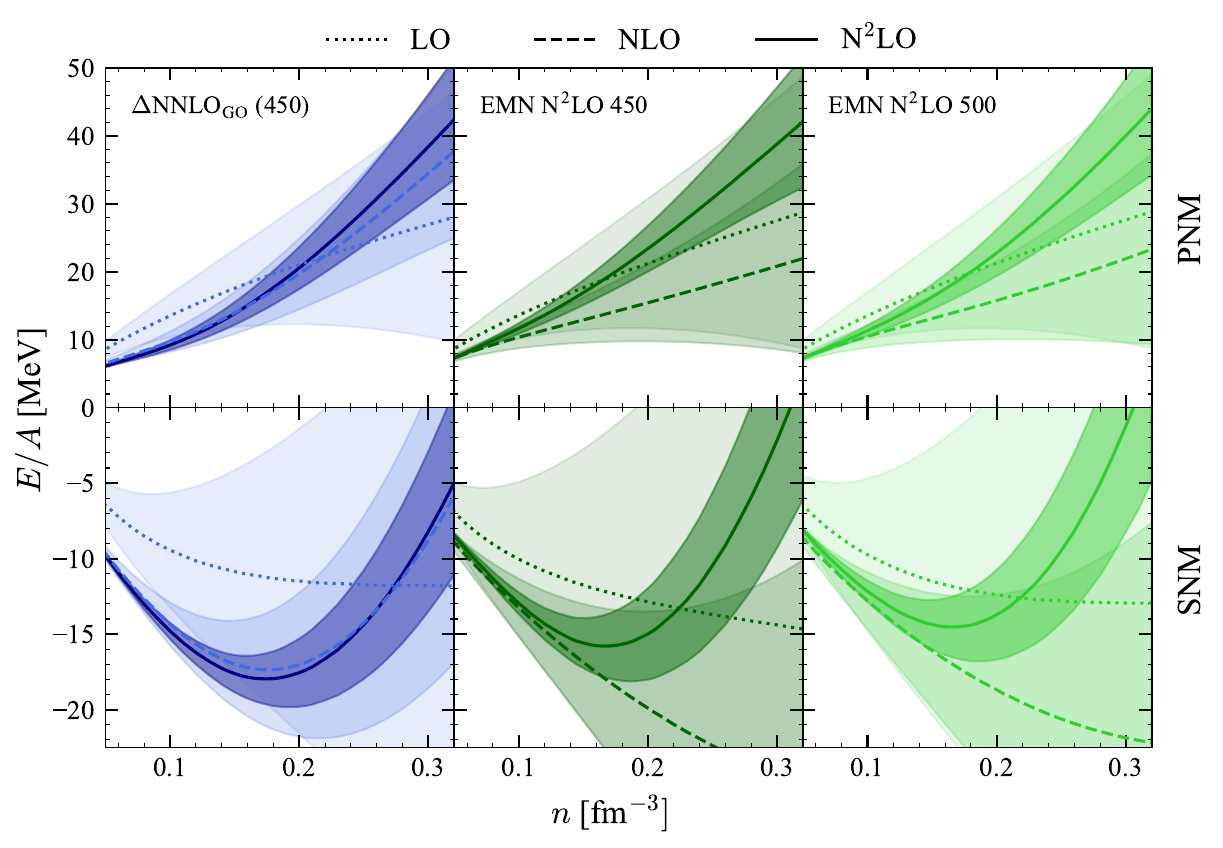}
    \caption{Energy per particle of neutron matter (top panels) and symmetric nuclear matter (lower panels) at different orders of the chiral expansion for the three different interactions, for which order-by-order potentials are available. All results are based on MBPT(3) calculations. Solid lines refer to results at N$^2$LO, dashed lines to NLO, and dotted lines to LO. The corresponding $68 \%$ confidence intervals based on GP-Bayesian uncertainties from the chiral EFT truncation (following Refs.~\cite{Drischler2020PRL,Drischler2020}) are indicated by dark bands at N$^2$LO, medium-dark bands at NLO, and light bands at LO.}
    \label{fig:EOS_chiral_uncertainties}
\end{figure*}

For SNM (see lower rows in Fig.~\ref{fig:EOS_MBPT}) we observe similar trends. Again, the smallest MBPT(2) correction at saturation density is found for the SRG-evolved interactions with $E^{(2)} = -2.6 \mev$ for the 1.8/2.0 (EM) interaction, whereas the MBPT(2) correction for the N$^2$LO EMN 500 interaction is about $E^{(2)} = -15.1 \mev$. At third order, we again observe a significant reduction in magnitude of the corrections for all employed interactions. While the overall contribution is again lowest for the 1.8/2.0 (EM) interaction with $E^{(3)} = 0.1 \mev$, we observe a particularly strong reduction for the $\Delta$-full interactions. It is striking that the N$^2$LO EMN 450 interaction shows a larger third-order correction over the whole range of densities compared to N$^2$LO EMN 500. This is likely due to the larger $c_D$ term for the $\Lambda = 450 \mev$ case. At $n_0$ we find $E^{(3)} = -2.1 \mev$ for N$^2$LO EMN 450 and $E^{(3)} = -0.8 \mev$ for N$^2$LO EMN 500, consistent with previous calculations~\cite{Drischler2020}. While the 1.8/2.0 (EM) and 1.8/2.0 (EM7.5) interactions by construction yield identical results for PNM, for SNM the larger $c_D$ for EM7.5 results in an increased 3N HF contribution from $E^{(1)}_{3N} = 7.6 \mev$ to $E^{(1)}_{3N} = 9.9 \mev$. This larger HF contribution partially cancels at second order, due to more attractive contributions in the second-order 3N diagrams. Hence, the shift of the LEC $c_D$ to larger values tends to make the interaction less perturbative, but improves on the other hand the agreement with empirical saturation properties of nuclear matter compared to the 1.8/2.0 (EM) interaction. In summary, for all studied interactions we find a systematic convergence pattern of the MBPT expansion and the many-body uncertainties are on the order of at most a few MeV in SNM at the highest density $2 n_0$, based on the size of the contributions at third order.

Compared to the MBPT uncertainties, the truncation uncertainties from the chiral EFT expansion of the interactions turn out to be significantly larger. To assess the truncation uncertainties, we need results at different orders. This is possible for the N$^2$LO EMN 450 and N$^2$LO EMN 500 interactions, where also leading order (LO) and NLO interactions are available, as well as for the $\Delta$-full interactions with $\Lambda = 450 \mev$. For the other $\Delta$NNLO$_\text{GO}$\,(394) interaction, no other orders are available. The low-resolution interactions 1.8/2.0 (EM), 1.8/2.0 (EM7.5), and 1.8/2.0 (sim7.5) have only been derived at specific orders in the chiral expansion and also include an SRG evolution of the NN contributions. This makes it difficult to assess the uncertainties from neglected higher-order interaction contributions.

For the determination of the EFT uncertainties for nuclear matter, we follow Refs.~\cite{Drischler2020PRL,Drischler2020} and employ a Gaussian process (GP) emulator to quantify the EFT truncation uncertainties. This assumes the energy per particle $E/A$ can be expanded in a power series,
\begin{equation}
\frac{E}{A}(n) = Y_\mathrm{ref}(n) \sum^\infty_{k=0} c_k(n) \, Q^k \,,
\label{eq:chiral_exp} 
\end{equation}
where the terms $k$ include the order-by-order contributions in the chiral expansion with the natural-size expansion coefficients $c_k(n)$. The expansion is in powers of $Q = p/\Lambda_b$, where $p$ is a typical momentum of the system and $\Lambda_b$ is the EFT breakdown scale. $Y_\mathrm{ref}(n)$ denotes the reference scale. In this work, we use the Fermi momentum $p = k_{\text{F}} = (6 \pi^2 n /g)^{1/3}$, with denegeracy $g=2$ ($g=4$) for PNM (SNM), as typical scale, and $\Lambda_b = 600 \mev$~\cite{Drischler2020PRL,Drischler2020}. The reference scale is taken to be $Y_\mathrm{ref}(n)= (n/n_0)^{2/3} \cdot 16 \mev$. We extract the coefficients $c_k$ from all orders up to N$^2$LO ($k = 0, 2, 3$) from our calculations and assume that the higher-order coefficients are normally distributed and of natural size. For more details on the resulting GP-Bayesian uncertainties see Refs.~\cite{Drischler2020PRL,Drischler2020}.

The resulting EFT truncation uncertainties at different orders for PNM and SNM are shown in Fig.~\ref{fig:EOS_chiral_uncertainties} for the different interactions, for which order-by-order potentials are available. Overall, we find that the N$^2$LO uncertainty bands for the N$^2$LO EMN 450 and N$^2$LO EMN 500 interactions as well as the $\Delta$-full interaction $\Delta$NNLO$_\text{GO}$\,(450) are quite similar, with an uncertainty of $\Delta E/A \pm (6-8) \mev$ in SNM and $\pm (9-10) \mev$ in PNM at the $68\%$ confidence level at $2 n_0$. Therefore, we do not find a significant reduction of the EFT truncation uncertainty in going from $\Delta$-less to $\Delta$-full interactions. However, what is more systematic for $\Delta$-full interactions is the improved convergence pattern in going from NLO to N$^2$LO (also Ref.~\cite{Ekst17deltasat}), because at NLO 3N interactions already enter and provide the necessary repulsion for saturation in SNM and lead to a stiffer EOS of PNM.

Finally, we can compare the different theoretical uncertainties and identify the dominant contributions. For the three interactions with order-by-order results, only the N$^2$LO EMN 450 interaction has a MBPT(3) correction that is similar to the EFT uncertainties around saturation density [for SNM at $n_0$ the MBPT(3) correction even slightly exceeds the EFT uncertainty at N$^2$LO]. To provide a combined EFT and many-body uncertainty, the inclusion of the MBPT truncation uncertainty will thus be important for future work. At higher densities, the EFT uncertainties of $\pm 8.2 \mev$ at $2 n_0$ are then large compared to the MBPT(3) corrections of $E^{(3)} = -3.3 \mev$. Compared to this interaction, the N$^2$LO EMN 500 interaction shows an improved MBPT convergence (as discussed above). As a consequence, the EFT uncertainty is largest over all densities for both PNM and SNM, while the estimated MBPT uncertainties based on the MBPT(3) corrections are still not entirely negligible. For the $\Delta$NNLO$_\text{GO}$\,(450) interaction, the EFT uncertainty dominates over the MBPT(3) correction. At $n_0$ we find EFT uncertainties of $\pm 1.9 \mev$ and $\pm 1.4 \mev$ for PNM and SNM, respectively, while at $2 n_0$ they are $\pm 8.9 \mev$ and $\pm 6.1 \mev$. These are all significantly larger than the corresponding MBPT(3) corrections. Thus, overall the Hamiltonian uncertainties dominate, but for a comprehensive determination of the total theoretical uncertainties, a framework that allows a joint analysis of chiral and many-body uncertainties is needed. 

\subsubsection{Equation of state parameters}

\begin{table}[t]
\begin{tabular}{L{2.3cm}|L{0.4cm}|L{1.0cm}L{1.0cm}L{0.8cm}|L{1.0cm}L{1.0cm}}
Interaction & $i$ & $n_\textrm{sat}$ & $E_{\textrm{sat}}$  & $K$ & $E_{\textrm{sym}}$ & $L$ \\
\hline
N$^2$LO EMN 450 & 2 & 0.145 & $-13.8$ & 244 & 31.6 & 49.3 \\
& 3 & 0.167 & $-15.8$ & 352 & 33.9 & 63.6 \\
\hline
N$^2$LO EMN 500 & 2 & 0.153 & $-13.8$ & 232 & 30.7 & 52.9  \\
& 3 & 0.165 & $-14.5$ & 323 & 32.0 & 60.7  \\
\hline
1.8/2.0 (EM) & 2 & 0.189 & $-17.1$ & 240 & 32.1 & 50.7  \\
& 3 & 0.191 & $-17.2$ & 251 & 32.2 & 51.5  \\ 
\hline
1.8/2.0 (EM7.5) & 2 & 0.162 & $-16.7$ & 270 & 32.1 & 42.1  \\
& 3 & 0.171 & $-17.6$ & 317 & 32.9 & 47.1  \\
\hline
1.8/2.0 (sim7.5) & 2 & 0.164 & $-16.9$ & 276 & 31.2 & 41.2  \\
& 3 & 0.172 & $-17.5$ & 320 & 32.0 & 46.7  \\
\hline
$\Delta$NLO$_{\textrm{GO}}\,(450)$ & 2 & 0.170 & $-17.0$ & 282 & 31.7 & 55.9  \\
& 3 & 0.176 & $-17.4$& 318 & 32.2 & 58.9  \\
\hline
$\Delta$NNLO$_{\textrm{GO}}\,(450)$ & 2 & 0.167 & $-17.6$ & 295 & 32.1 & 64.1  \\
& 3 & 0.174 & $-18.0$& 338 & 32.9 & 67.3  \\
\hline
$\Delta$NNLO$_{\textrm{GO}}\,(394)$ & 2 & 0.159 & $-16.6$& 250 & 32.2 & 56.8  \\
& 3 & 0.165 & $-16.9$& 282 & 32.7 & 58.9  \\
\end{tabular}
\caption{Equation of state parameters for the employed interactions at MBPT(2) and MBPT(3) ($i=2$ and 3, respectively). Note that the incompressibility $K$ of SNM is given for the calculated saturation density $n_\textrm{sat}$, while the symmetry energy $E_{\textrm{sym}}$ and the $L$ parameter are evaluated at $n_0 = 0.16 \fmiq$. $n_\textrm{sat}$ is given in fm$^{-3}$; all other quantities are in MeV.
\label{tab:EOS_Parameters}}
\end{table}

The results of our calculations allow the extraction of characteristic parameters of the EOS for the different interactions. For SNM, we use the expansion of the energy per particle around the saturation density:
\begin{equation} 
    \frac{E}{A}(n) = E_{\textrm{sat}} + \frac{1}{2} \, K \, \left( \frac{n - n_\textrm{sat}}{3 n_\textrm{sat}} \right)^2 + \ldots \,.
    \label{eq:EOS_expansion}
\end{equation}
Fitting Eq.~\eqref{eq:EOS_expansion} to our SNM results around the calculated saturation density for the different interactions yields results for the calculated $n_\textrm{sat}$, the saturation energy $E_\textrm{sat}$, and the incompressibility $K$. Note that we use $n_\textrm{sat}$ for the saturation density of the calculation, and $n_0 = 0.16 \fmiq$ for the canonical reference value used, e.g., in astrophysical contexts.

For the symmetry energy $S(n)$, we extract this from the difference of energy per particle for PNM and SNM:
\begin{equation}
S(n) = \frac{E_{\mathrm{PNM}}}{N}(n) - \frac{E_{\mathrm{SNM}}}{A}(n) \,.
\end{equation}
This is a particular choice when only results for PNM and SNM are available. We expand the density dependence of $S(n)$ around the canonical value $n_0 = 0.16 \fmiq$:
\begin{equation}
    S(n) = E_{\mathrm{sym}} + L \frac{n - n_0}{3 n_0} + \ldots \,,
\end{equation}
with the symmetry energy $E_{\textrm{sym}}$ at $n_0$ and the $L$ parameter, defined by $L = L(n_0) = 3 n_0 \left. \frac{\partial S(n)}{\partial n} \right \vert_{n=n_0}$. Our results for $E_{\textrm{sym}}$ and $L$ are then directly taken from $S(n_0)$ and from the numerical derivative at $n_0$, respectively. We believe this is a useful input for astrophysics, because one usually compares $E_{\textrm{sym}}$ and $L$ at a fixed density $n_0$. If we instead evaluate the $L$ parameter at the calculated saturation density $n_\textrm{sat}$, the obtained $L$ parameter should not be compared to other studies, because $L$ approximately scales linearly with $n_\textrm{sat}$.

Table~\ref{tab:EOS_Parameters} gives our results for the equation of state parameters for the different interactions at MBPT(2) and MBPT(3). The results for the calculated saturation point follow the trend discussed for Fig.~\ref{fig:EOS_SNM}, where interactions that have been constrained by nuclei and/or nuclear matter naturally provide a better description of saturation. For the incompressibility $K$, we obtain a range between $251 \mev \leqslant K \leqslant 352 \mev$ at MBPT(3). This broad range is consistent with a survey from nuclear energy-density functionals~\cite{Dutra:2012mb}. On the other hand, we find that the symmetry energy $E_{\text{sym}}$ is rather tightly constrained to the range $32.0 \mev \leqslant E_{\text{sym}} \leqslant 33.9 \mev$ for the different interactions at MBPT(3). Naturally, the $L$ parameter as a derivative shows a larger variation of $46.7 \mev \leqslant L \leqslant 67.3 \mev$. Both $E_{\text{sym}}$ and $L$ are consistent with recent constraints (see, e.g., Refs.~\cite{Essick2021,Huth:2020ozf,Lim:2023dbk}). Comparing the different interactions, one observes that the SRG-evolved interactions 1.8/2.0 yield the lowest $L$ parameter, reflecting the softer behavior of the corresponding PNM EOS (see Sec.~\ref{chap:EOS_Results}). Finally, we emphasize that the given ranges only represent an estimate of the Hamiltonian uncertainty. A full uncertainty quantification would entail including correlated EFT uncertainties and many-body uncertainties, which we leave to future work.

\section{Fermi liquid parameters}
\label{chap:FLT}

\subsection{Fermi liquid theory and calculational details}

The Fermi liquid theory (FLT) approach to nucleonic matter based on an MBPT expansion with NN and 3N interactions has been discussed in Refs.~\cite{Schwenk:2002fq,Schwenk:2003bc,Friman:2012ft,Holt:2011,Holt:2013}. Here, we include all diagrams up to second order in MBPT. The key quantity of interest in FLT is the quasiparticle interaction $\mathcal{F} (\mathbf{p}_1 \sigma_1 \tau_1,\mathbf{p}_2 \sigma_2 \tau_2)$, which is given by the second functional differentiation of the energy density $\mathcal{E} = E/V$ of the system with respect to the quasiparticle occupation numbers
\begin{equation}
    \mathcal{F} (\textbf{p}_1 \sigma_1 \tau_1,\textbf{p}_2 \sigma_2 \tau_2) = \frac{\delta^2 \mathcal{E}}{\delta n_{\textbf{p}_1\sigma_1\tau_1}\delta n_{\textbf{p}_2\sigma_2\tau_2}} \,.
    \label{eq:quasi-energy}
\end{equation}
Here $\textbf{p}_i$ is the quasiparticle momentum (for well defined quasiparticles $\textbf{p}_i$ are close to the Fermi surface), $\sigma_i = \pm 1/2$ are the spin projections, and $\tau_i = p,n$ the isospin.

In this work, we focus on Fermi liquid parameters for PNM. Focusing only on the central components, the quasiparticle interaction can be decomposed into
\begin{equation}
    \mathcal{F} (\textbf{p}_1 \sigma_1,\textbf{p}_2 \sigma_2) = f(\textbf{p}_1,\textbf{p}_2) + g(\textbf{p}_1,\textbf{p}_2) \, \boldsymbol{\sigma}_1 \cdot \boldsymbol{\sigma}_2 \,.
    \label{eq:quasipart_int}
\end{equation}
Restricting the quasiparticle momenta $\mathbf{p}_1$ and $\mathbf{p}_2$ to the Fermi surface, the components of the quasiparticle interaction can be expanded in Legendre polynomials
\begin{equation}
    f(\textbf{p}_1,\textbf{p}_2) = \sum_{l=0}^\infty f_l \, P_l(\cos\theta) \,,
\end{equation}
and similarly for $g(\textbf{p}_1,\textbf{p}_2)$. Here $\theta$ is the angle between $\textbf{p}_1$ and $\textbf{p}_2$ and the coefficients $f_l$ are the Landau parameters. From these the dimensionless Landau parameters are given by
\begin{equation}
    F_l = N(0) \, f_l \,,
\end{equation}
with the density of states at the Fermi surface $N(0) = m^* k_\text{F}/\pi^2$, where $k_\text{F}$ is the Fermi momentum and the effective mass $m^*$ has to be calculated self-consistently from $m^*/m = 1 + F_1/3$. The Landau parameters $g_l$ and $G_l$ are defined accordingly.

\begin{figure*}
    \centering
    \includegraphics[width=0.845\linewidth]{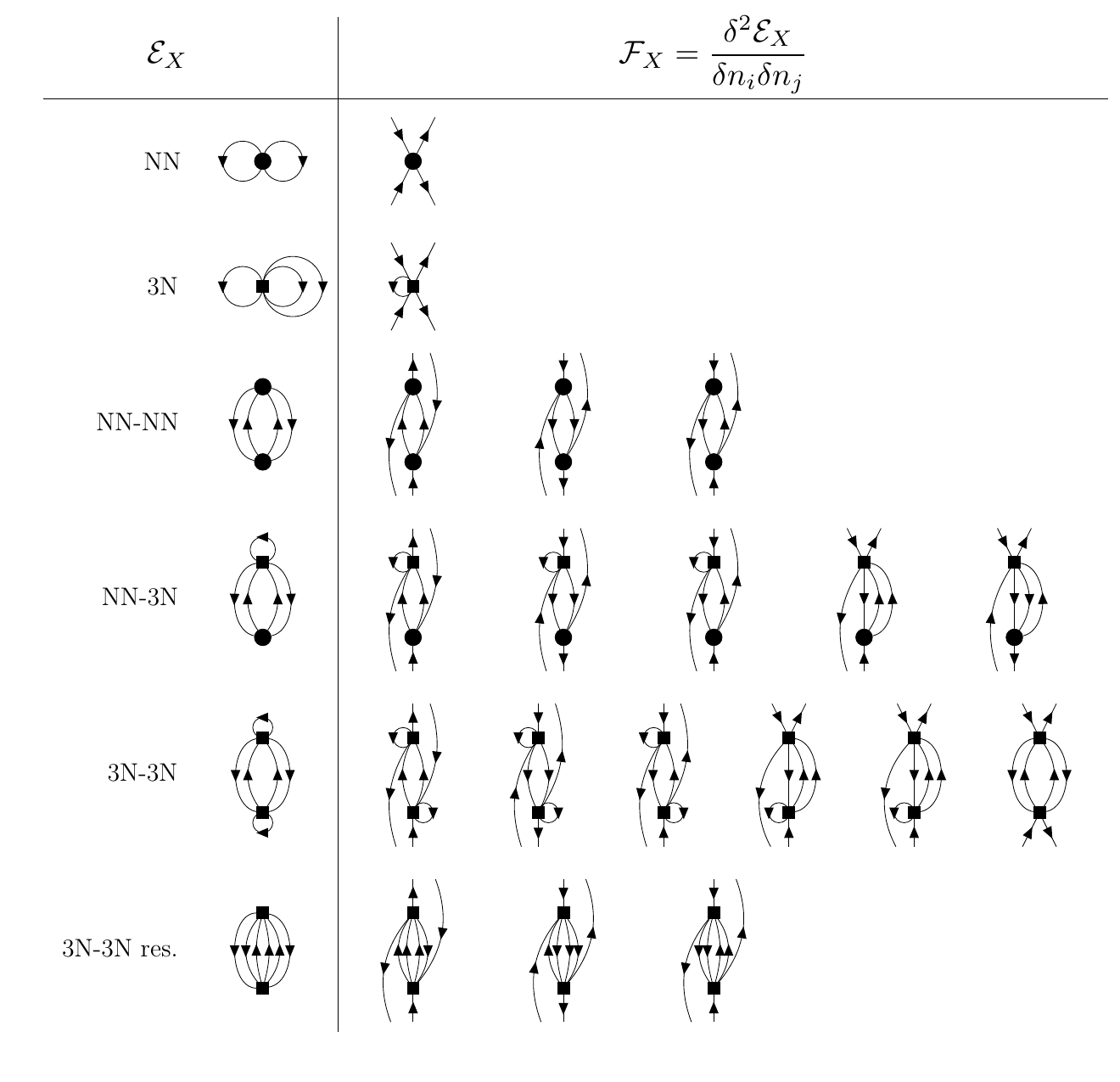}
    \caption{Hartree-Fock and second-order diagrams that contribute to the quasiparticle interaction. The left column shows the  diagrams with NN and 3N vertices contributing to the energy density $\mathcal{E}$. The diagrams on the right show the resulting diagrammatic contributions to the quasiparticle interaction~$\mathcal{F}$. Upward (downward) lines represent particles (holes).}
    \label{fig:flt_diagrams}
\end{figure*}

In this work, we focus on the spin-independent quasiparticle interaction $f(\textbf{p}_1,\textbf{p}_2)$, as this is most relevant for the equation of state. This part can be obtained from $\mathcal{F}$ by tracing over spin $f = {\rm Tr}_{\sigma_1,\sigma_2} \mathcal{F}/4$. Given a MBPT contribution to the energy density $\mathcal{E}^{(n)}$, we can obtain the corresponding contribution to the quasiparticle interaction $\mathcal{F}^{(n)}$ by functional differentiation:
\begin{equation}
    \mathcal{F}^{(n)}(1,2) = \mathcal{F}^{(n)}(\textbf{p}_1 \sigma_1, \textbf{p}_2 \sigma_2) = \frac{\delta^2 \mathcal{E}^{(n)}}{\delta n_{\textbf{p}_1\sigma_1}\delta n_{\textbf{p}_2\sigma_2}} \,.
\end{equation}
Starting from NN and 3N interactions, there are two energy contributions at the Hartree-Fock level and five at second order in MBPT. The quasiparticle interaction at these orders is then given by
\begin{align}
    \mathcal{F}^{(1)}(1,2) &= \frac{\delta^2}{\delta n_{1}\delta n_{2}}\left(\mathcal{E}^{(1)}_{\text{NN}} + \mathcal{E}^{(1)}_{\text{3N}}\right) \,, \\
    \mathcal{F}^{(2)}(1,2) &= \frac{\delta^2}{\delta n_{1}\delta n_{2}} \left(\mathcal{E}^{(2)}_{\text{NN-NN}} + \mathcal{E}^{(2)}_{\text{NN-3N}}+\mathcal{E}^{(2)}_{\text{3N-NN}} \right. \nonumber \\
    &\quad \left. + \mathcal{E}^{(2)}_{\text{3N-3N}}+\mathcal{E}^{(2)}_{\text{3N-3N res.}}\right) \,.
\end{align}

The Hartree-Fock and second-order contributions used for the calculation of the quasiparticle interactions are depicted in Fig.~\ref{fig:flt_diagrams}. At the Hartree-Fock level we have
\begin{align}
    \mathcal{E}^{(1)}_{\text{NN}} &= \frac{1}{2}\sum_{ij}\vnn{i}{j}{i}{j} \,, \\
    \mathcal{F}^{(1)}_{\text{NN}}(1,2) &= \vnn{1}{2}{1}{2} \,, \\
    \mathcal{E}^{(1)}_{\text{3N}} &= \frac{1}{6} \sum_{ijk} \vtn{i}{j}{k}{i}{j}{k} \,, \\
    \mathcal{F}^{(1)}_{\text{3N}}(1,2) &= \sum_{k}\vtn{1}{2}{k}{1}{2}{k} \,,
\end{align}
where we have used the notation that sums over roman indices correspond to sums over hole states, and sums over greek indices to particle states:
\begin{equation}
    \sum_i =\mathrm{Tr}_{\sigma_i} \int\frac{d\textbf{k}_i}{(2\pi)^3} \, n_{\textbf{k}_i} \quad \text{and} \quad \sum_\alpha = \mathrm{Tr}_{\sigma_\alpha} \int\frac{d\textbf{k}_\alpha}{(2\pi)^3}(1-n_{\textbf{k}_\alpha}) \,.
\end{equation}
$n_{\textbf{k}} = \theta(|\textbf{k}|-k_{\rm F})$ is the Fermi-Dirac distribution function given by the theta function at zero temperature.

The calculation of the second-order contributions follows the same strategy. Starting from the NN-NN energy diagram one obtains
\begin{align}
    \mathcal{E}^{(2)}_{\text{NN-NN}} &= \frac{1}{4}\sum_{\substack{ij\\ \alpha\beta}} \frac{\bigl|\vnn{i}{j}{\alpha}{\beta} \, \bigr|^2}{D^{ij}_{\alpha\beta}} \,, \\
    \mathcal{F}^{(2),\text{I}}_{\text{NN-NN}} &= \frac{1}{2}\sum_{\alpha\beta} \frac{\bigl|\vnn{1}{2}{\alpha}{\beta} \, \bigr|^2}{D^{12}_{\alpha\beta}} \,, \\
    \mathcal{F}^{(2),\text{II}}_{\text{NN-NN}} &= \frac{1}{2}\sum_{ij} \frac{\bigl|\vnn{i}{j}{1}{2} \, \bigr|^2}{D^{ij}_{12}} \,, \\
    \mathcal{F}^{(2),\text{III}}_{\text{NN-NN}} &= -2\sum_{{j\beta}} \frac{\bigl|\vnn{1}{j}{2}{\beta} \, \bigr|^2}{D^{1j}_{2\beta}} \,,
\end{align}
where the roman superscript labels the diagrammatic contribution to the quasiparticle interaction in Fig.~\ref{fig:flt_diagrams} (counting from left to right). Moreover, we use the compact notation for the energy denominator $D^{ij\ldots}_{\alpha\beta\ldots} = \varepsilon_i+\varepsilon_j+ \ldots -\varepsilon_\alpha-\varepsilon_\beta-\ldots$. The single-particle energies are calculated at the HF level
\begin{equation}
    \varepsilon(\textbf{p}) = \frac{\textbf{p}^2}{2m} + \sum_j\vnn{\textbf{p}}{j}{\textbf{p}}{j} + \frac{1}{2}\sum_{jk}\vtn{\textbf{p}}{j}{k}{\textbf{p}}{j}{k} \,,
\end{equation}
for both particle and hole states, and $\varepsilon(\textbf{p})$ is the same for both spin states.

When 3N interactions are included, there are three additional topologies for the second-order energy diagrams: two diagrams where one of the NN vertices is replaced by a 3N vertex with one line closed (i.e., a normal-ordered two-body contribution from the 3N interaction), one diagram where both NN vertices are replaced by normal-ordered 3N vertices, and a residual (``res.'') 3N-3N diagram. Note that in Fig.~\ref{fig:flt_diagrams} we only show one of the NN-3N diagrams, as the other contributions follow by interchanging the NN and normal-ordered 3N vertex. Since the quasiparticle interaction is symmetric under exchanging 1 and 2, these contributions simply double the NN-3N one. For the NN-3N diagrams, we have
\begin{align}
    \mathcal{E}^{(2)}_{\text{NN-3N}} &= \frac{1}{4}\sum_{\substack{ijk\\\alpha\beta}} \frac{\vnn{i}{j}{\alpha}{\beta}\vtn{\alpha}{\beta}{k}{i}{j}{k}}{D^{ij}_{\alpha\beta}} \,, \\
    \mathcal{F}^{(2),\text{I}}_{\text{NN-3N}}(1,2) &=  \frac{1}{2}\sum_{\substack{\alpha\beta k}} \frac{\vnn{1}{2}{\alpha}{\beta}\vtn{\alpha}{\beta}{k}{1}{2}{k}}{D^{12}_{\alpha\beta}}\,, \\
    \mathcal{F}^{(2),\text{II}}_{\text{NN-3N}}(1,2) &=  \frac{1}{2}\sum_{\substack{ijk}} \frac{\vnn{i}{j}{1}{2}\vtn{1}{2}{k}{i}{j}{k}}{D^{ij}_{12}} \,, \\
    \mathcal{F}^{(2),\text{III}}_{\text{NN-3N}}(1,2) &= -2\sum_{\substack{jk\beta}} \frac{\vnn{1}{j}{2}{\beta}\vtn{2}{\beta}{k}{1}{j}{k}}{D^{1j}_{2\beta}} \,, \\
    \mathcal{F}^{(2),\text{IV}}_{\text{NN-3N}}(1,2) &=\sum_{{j\alpha\beta}} \frac{\vnn{1}{j}{\alpha}{\beta}\vtn{\alpha}{\beta}{2}{1}{j}{2}}{D^{1j}_{\alpha\beta}} \,, \\
    \mathcal{F}^{(2),\text{V}}_{\text{NN-3N}}(1,2) &= -\sum_{\substack{ij\beta}} \frac{\vnn{i}{j}{1}{\beta}\vtn{1}{\beta}{2}{i}{j}{2}}{D^{ij}_{1\beta}} \,.
\end{align}
Note that these have to be doubled for the total NN-3N plus 3N-NN contributions.

For the 3N-3N second-order contributions, there are two different energy diagrams. One in which the 3N vertices enter as normal-ordered two-body interactions, and one from the residual 3N-3N contributions, which we label ``3N-3N res.''. For the former one, we have
\begin{align}
    \mathcal{E}^{(2)}_{\text{3N-3N}} &= \frac{1}{4}\sum_{\substack{ijkl\\ \alpha\beta}} \frac{\vtn{\mathnormal{i}}{\mathnormal{j}}{\mathnormal{k}}{\alpha}{\beta}{k}\vtn{\alpha}{\beta}{\mathnormal{l}}{i}{j}{\mathnormal{l}}}{D^{ij}_{\alpha\beta}} \,, \\
    \mathcal{F}^{(2),\text{I}}_{\text{3N-3N}} &= \frac{1}{2} \sum_{\substack{kl\alpha\beta}} \frac{\vtn{1}{2}{\mathnormal{k}}{\alpha}{\beta}{k}\vtn{\alpha}{\beta}{\mathnormal{l}}{1}{2}{\mathnormal{l}}}{D^{12}_{\alpha\beta}} \,, \\
    \mathcal{F}^{(2),\text{II}}_{\text{3N-3N}} &= \frac{1}{2}\sum_{\substack{ijkl}} \frac{\vtn{\mathnormal{i}}{\mathnormal{j}}{\mathnormal{k}}{1}{2}{k}\vtn{1}{2}{\mathnormal{l}}{i}{j}{\mathnormal{l}}}{D^{ij}_{12}} \,, \\
    \mathcal{F}^{(2),\text{III}}_{\text{3N-3N}} &= -2\sum_{\substack{jkl\beta}} \frac{\vtn{1}{\mathnormal{j}}{\mathnormal{k}}{2}{\beta}{k}\vtn{2}{\beta}{\mathnormal{l}}{1}{j}{\mathnormal{l}}}{D^{1j}_{2\beta}} \,, \\
    \mathcal{F}^{(2),\text{IV}}_{\text{3N-3N}} &= 2\sum_{\substack{jl\alpha\beta}} \frac{\vtn{1}{\mathnormal{j}}{2}{\alpha}{\beta}{2}\vtn{\alpha}{\beta}{\mathnormal{l}}{1}{j}{\mathnormal{l}}}{D^{1j}_{\alpha\beta}} \,, \\
    \mathcal{F}^{(2),\text{V}}_{\text{3N-3N}} &=  -2\sum_{\substack{ijl\beta}} \frac{\vtn{\mathnormal{i}}{\mathnormal{j}}{2}{1}{\beta}{2}\vtn{1}{\beta}{\mathnormal{l}}{i}{j}{\mathnormal{l}}}{D^{ij}_{1\beta}} \,, \\
    \mathcal{F}^{(2),\text{VI}}_{\text{3N-3N}} &= \frac{1}{2}\sum_{\substack{ij\alpha\beta}} \frac{\vtn{\mathnormal{i}}{\mathnormal{j}}{1}{\alpha}{\beta}{1}\vtn{\alpha}{\beta}{2}{i}{j}{2}}{D^{ij}_{\alpha\beta}} \,.
\end{align}
For the residual 3N-3N diagram, we have
\begin{align}
    \mathcal{E}^{(2)}_{\text{3N-3N res.}} &= \frac{1}{36}\sum_{\substack{ijk\\ \alpha\beta\gamma}} \frac{\vtn{\mathnormal{i}}{\mathnormal{j}}{\mathnormal{k}}{\alpha}{\beta}{\gamma}\vtn{\alpha}{\beta}{\gamma}{i}{j}{k}}{D^{ijk}_{\alpha\beta\gamma}} \,, \\
    \mathcal{F}^{(2),\text{I}}_{\text{3N-3N res.}} &= \frac{1}{6}\sum_{\substack{k\alpha\beta\gamma}} \frac{\vtn{1}{2}{\mathnormal{k}}{\alpha}{\beta}{\gamma}\vtn{\alpha}{\beta}{\gamma}{1}{2}{k}}{D^{12k}_{\alpha\beta\gamma}} \,, \\
    \mathcal{F}^{(2),\text{II}}_{\text{3N-3N res.}} &= \frac{1}{6}\sum_{\substack{ijk\gamma}} \frac{\vtn{\mathnormal{i}}{\mathnormal{j}}{\mathnormal{k}}{1}{2}{\gamma}\vtn{1}{2}{\gamma}{i}{j}{k}}{D^{ijk}_{12\gamma}} \,, \\
    \mathcal{F}^{(2),\text{III}}_{\text{3N-3N res.}} &= -\frac{1}{2}\sum_{\substack{jk\beta\gamma}} \frac{\vtn{1}{\mathnormal{j}}{\mathnormal{k}}{2}{\beta}{\gamma}\vtn{2}{\beta}{\gamma}{1}{j}{k}}{D^{1jk}_{2\beta\gamma}} \,.
\end{align}

\subsection{Landau parameters}

\begin{figure}[t!]
    \centering
    \includegraphics[width=\columnwidth]{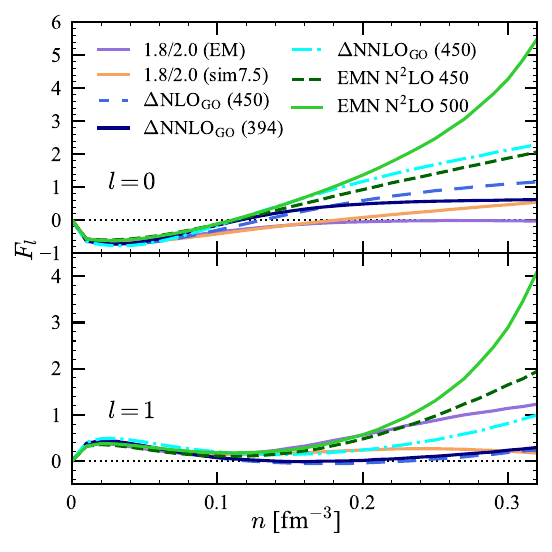}
    \caption{Landau parameters $F_0$ (upper panel) and $F_1$ (lower panel) for PNM as a function of density $n$ for the different interactions considered. Results are shown at the MBPT(2) level including all NN and 3N contributions fully.}
    \label{fig:F0F1-pot-spread}
\end{figure}

\begin{figure*}[t!]
    \centering
    \includegraphics[width=0.85\textwidth]{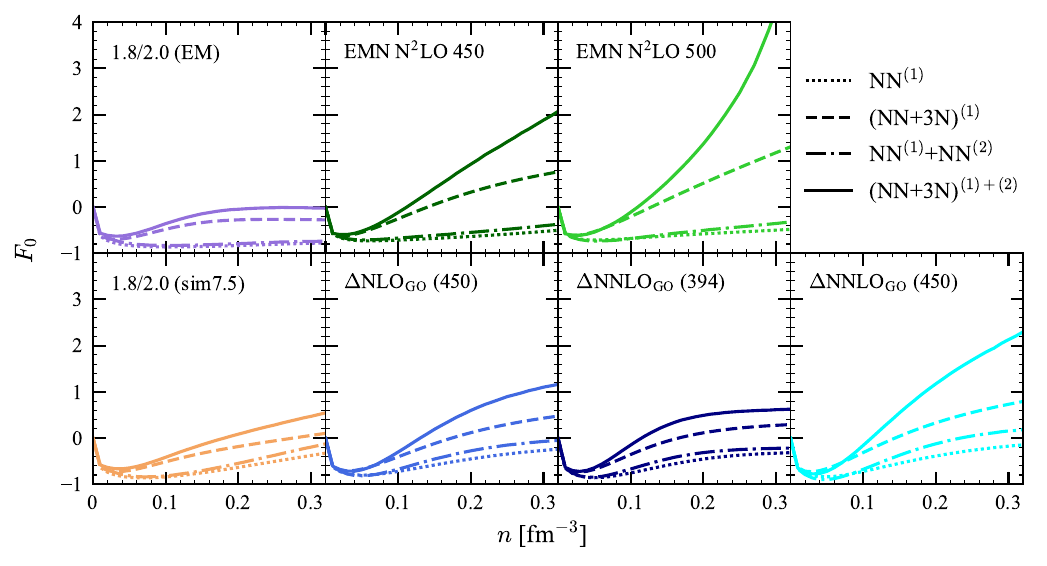}
    \caption{Landau parameter $F_0$ for PNM for different NN and 3N interactions at different orders in the MBPT expansion. Results are shown at the HF level, with dotted lines for the NN and dashed lines for the 3N contributions, as well as including MBPT(2) contributions, with dot-dashed lines for NN-only and solid lines for the full NN+3N results.}
    \label{fig:F0_MBPT_table}
\end{figure*}

\begin{figure*}[t!]
    \centering
    \includegraphics[width=0.85\textwidth]{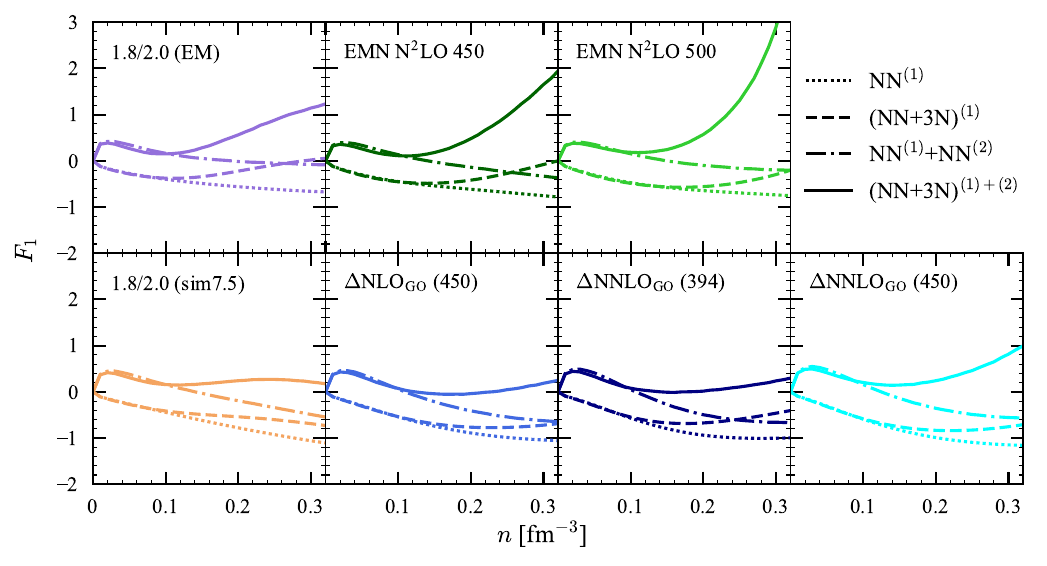}
    \caption{Same as Fig.~\ref{fig:F0_MBPT_table} but for the Landau parameter $F_1$.}
    \label{fig:F1_MBPT_table}
\end{figure*}

Our MBPT(2) results for the Landau parameters $F_l$ for $l=0,1$\ are shown in Fig.~\ref{fig:F0F1-pot-spread} for the different interactions considered. Note that our lowest density point calculated is for $n=0.01\fmiq$.\ For $F_0$, we observe a common attractive behavior up to densities around $0.1 \fmiq$ for all interactions. With the exception of the 1.8/2.0 (EM) interaction, all $F_0$ become repulsive above saturation density. This repulsion further increases to higher densities, except for the $\Delta$NNLO$_{\textrm{GO}}\,(394)$ interaction. Since $F_0$ is related to the speed of sound (see Sec.~\ref{chap:FLT_meffcs2}), larger values indicate a stiffer EOS. Similarly to $F_0$, we also find a common behavior for $F_1$, again up to densities around $0.1 \fmiq$, leading to an enhanced effective mass $m^*/m > 1$. We attribute these common behaviors at low densities for $F_0$ and $F_1$ due to being dominated by NN interactions. The enhancement of the effective mass is consistent with more sophisticated calculations that resum particle-hole contributions to the quasiparticle interactions~\cite{Schwenk:2002fq}. Furthermore, we observe a significant dependence of the results for $F_0$ and $F_1$ on the cutoff for densities beyond $1.5 n_0$. For instance, the Landau parameters become significantly more repulsive for the N$^2$LO EMN interaction with $\Lambda = 500 \mev$ compared to $\Lambda = 450 \mev$. 
For $F_1$, this directly translates to larger effective masses at higher densities. 

In Figs.~\ref{fig:F0_MBPT_table}~and~\ref{fig:F1_MBPT_table} the MBPT convergence and the effect of the different diagrams on the Landau parameters are shown. For $F_0$ in Fig.~\ref{fig:F0_MBPT_table}, the HF NN contribution is attractive for the entire density range for all interactions. The addition of the second-order NN contribution only moderately changes $F_0$, so that $F_0$ is still attractive overall, except for the $\Delta$NNLO$_{\textrm{GO}}\,(394)$ interaction, where the second-order NN contributions make $F_0$ slightly repulsive at higher densities. For all interactions, 3N forces provide the important repulsive contributions to $F_0$. The 3N repulsive effect is weakest for the low-resolution interactions $1.8/2.0$ and most prominent for the EMN N$^2$LO $450/500$ and $\Delta$NNLO$_{\textrm{GO}}\,(450)$ cases.

For $F_1$ in Fig.~\ref{fig:F1_MBPT_table}, we find that NN interactions at the HF level lead to a decreasing $F_1$ that is similar for all interactions up to $0.1 \fmiq$, where repulsive 3N contributions start to be important and increase $F_1$ with increasing density. The resulting $F_1$ at the HF level is nevertheless still $\lesssim 0$ (i.e., $m^*/m < 1$) up to $2 n_0$. Including second-order NN contributions leads to an enhancement of the effective mass relative to the HF NN value. This effect from the $\omega$-mass is due to the energy dependence of the neutron self-energy~\cite{HebelerSchwenk2010}. Including the full MBPT(2) contribution adds the repulsive 3N effects also at second order, so that $F_1$ becomes positive (i.e., $m^*/m > 1$) for all densities and interactions considered. This increasing repulsion is largest for the EMN N$^2$LO $450/500$ interactions. Finally, we note that both Landau parameters obey the Pomeranchuk stability criterion $F_l > -(2l+1)$~\cite{Pomeranchuk:1959}.

\subsection{Effective mass and speed of sound}
\label{chap:FLT_meffcs2}

As mentioned, the effective mass $m^*$ is related to the Landau parameter $F_1$ through $m^*/m = 1 + F_1/3$ and the speed of sound is given from $F_0$ and $m^*$ by~\cite{BaymPethick:1991}
\begin{equation}
    c_s^2 =\frac{1}{3} \frac{k_{\rm F}^2}{m^*\mu}(1+F_0) \,,
    \label{eq:cs}
\end{equation}
where $\mu$ denotes the chemical potential, which we determine self-consistently from $f_0$ and $f_1$ through~\cite{BaymPethick:1991}
\begin{equation}
    \frac{\partial\mu}{\partial n} = \frac{\pi^2}{m k_{\rm F}} + f_0 - \frac{1}{3}f_1 \,.
\end{equation}
We have checked that for $c_s^2$ this gives consistent results when comparing against $c_s^2 = \partial P/\partial \mathcal{E}$ obtained from the EOS calculations in Sec.~\ref{chap:EOS_Results} through thermodynamic derivatives.
We have checked that for $c_s^2$ this gives consistent results when comparing against $c_s^2 = \partial P/\partial \mathcal{E}$ obtained from the EOS calculations in Sec.~\ref{chap:EOS_Results} through thermodynamic derivatives.

\begin{figure}[t!]
    \centering
    \includegraphics[width=\columnwidth]{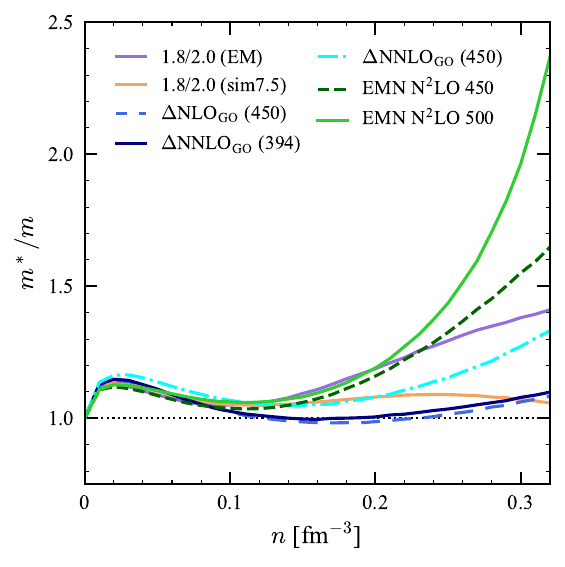}
    \caption{Effective mass $m^*/m$ in PNM as a function of density $n$ for the different interactions considered. Results are shown at the MBPT(2) level including all NN and 3N contributions fully.}
    \label{fig:eff_mass_pots}
\end{figure}

Figure~\ref{fig:eff_mass_pots} shows our MBPT(2) results for the effective mass for the different interactions. Since the results for $m^*$ closely resemble the behavior of $F_1$ shown in Fig~\ref{fig:F0F1-pot-spread}, we observe the same common behavior up to densities $0.1 \fmiq$, and more interaction uncertainties with increasing density. As discussed for $F_1$, due to the 3N contributions, the effective mass increases for larger densities with $m^*/m > 1$ for all interactions. Beyond saturation density, there is a significant increase in $m^*$ for the EMN N$^2$LO $450/500$ interactions, but also for the 1.8/2.0 (EM) and the $\Delta$NNLO$_{\textrm{GO}}\,(450)$ interactions. Again, we find a significant cutoff dependence for the EMN N$^2$LO and $\Delta$NNLO$_{\text{GO}}$ interactions. 

\begin{figure}[t!]
    \centering
    \includegraphics[width=\columnwidth]{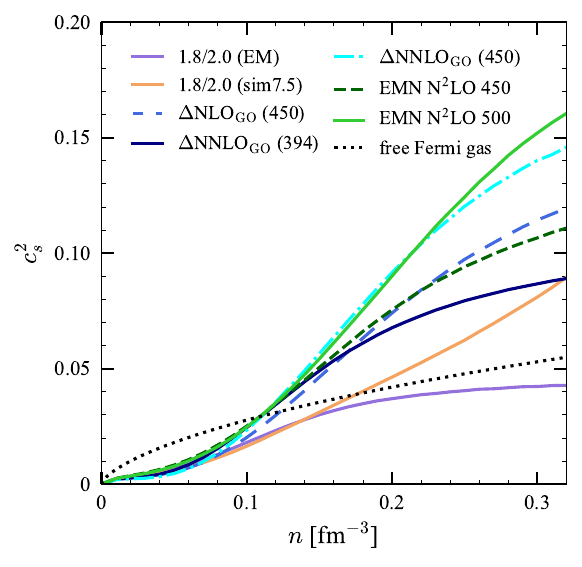}
    \caption{Same as Fig.~\ref{fig:eff_mass_pots} but for the speed of sound squared $c_s^2$.}
    \label{fig:cs_pots}
\end{figure}

Our MBPT(2) results for the speed of sound squared for the different interactions are given in Fig.~\ref{fig:cs_pots}. We find again a common behavior up to $0.1 \fmiq$. Beyond this density, the behavior depends again more sensitively on the interactions, as expected from $F_0$ and $F_1$. The low-resolution interactions 1.8/2.0 give a smaller $c_s^2$, whereas the other interactions can reach $c_s^2 \lesssim 0.15$ up to $2 n_0$. Note that the significant increase above the free Fermi gas value in Fig.~\ref{fig:cs_pots} is due to the repulsive 3N contributions.

\begin{figure*}[t]
    \centering
    \includegraphics[width=0.75\textwidth]{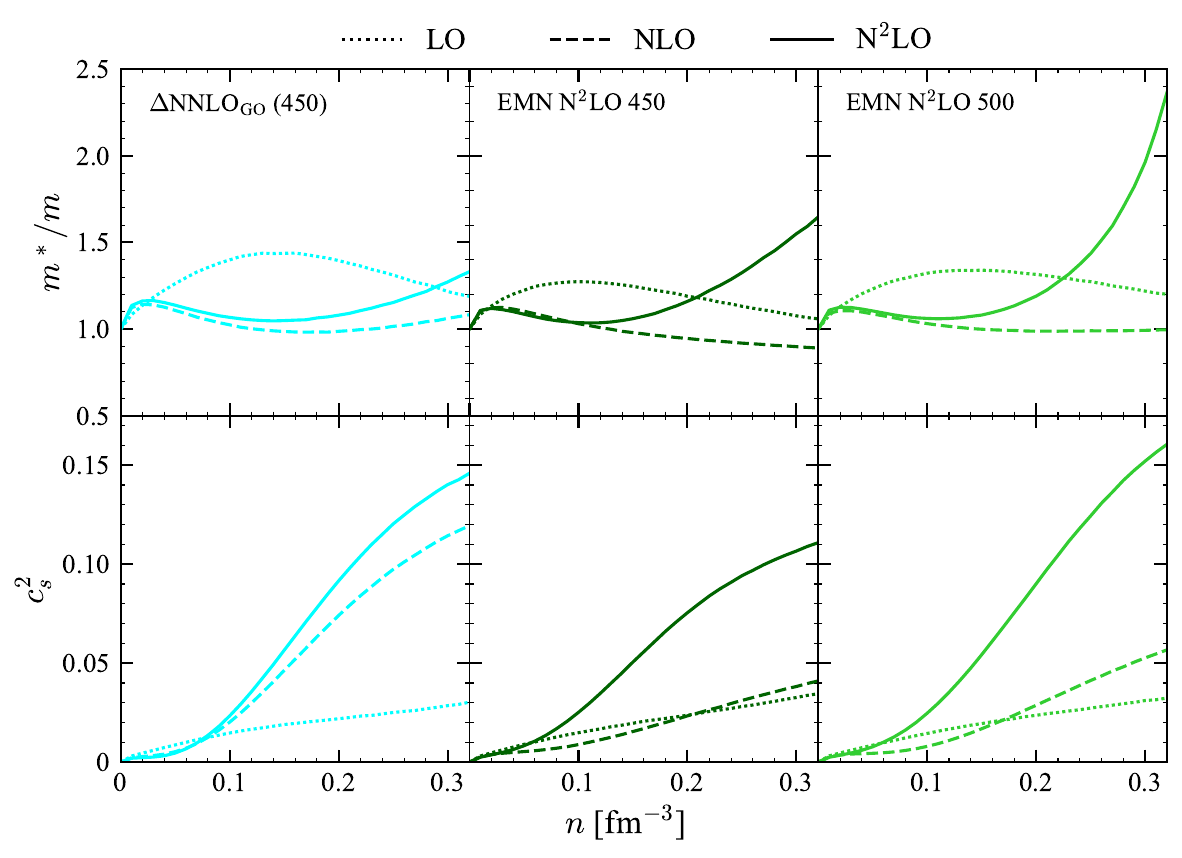}
    \caption{Effective mass $m^*/m$ (upper panels) and speed of sound squared $c_s^2$ (lower panels) in PNM at different orders of the chiral expansion for the three different interactions, for which order-by-order potentials are available. All results are based on MBPT(2) calculations. Solid lines refer to results at N$^2$LO, dashed lines to NLO, and dotted lines to LO.}
    \label{fig:mEff_EFT_order}
\end{figure*}

\begin{figure*}[t!]
    \centering
    \includegraphics[width=0.85\textwidth]{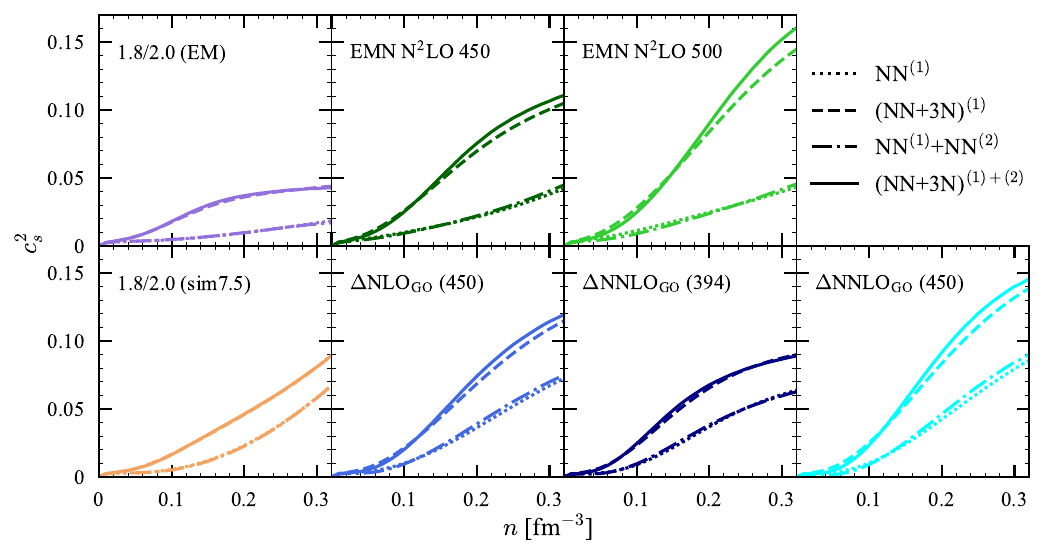}
    \caption{Same as Fig.~\ref{fig:F0_MBPT_table} but for the speed of sound squared $c_s^2$.}
    \label{fig:cs_MBPT_table}
\end{figure*}

Figure~\ref{fig:mEff_EFT_order} shows our MBPT(2) results for the effective mass and the speed of sound at different orders of the chiral expansion for the three different interactions, for which order-by-order potentials are available. At LO the behavior of the effective mass and the speed of sound is quite similar for all interactions. At NLO, both EMN 450/500 interactions remain similar, but the $\Delta$NLO$_{\textrm{GO}}\,(450)$ interaction leads to a larger speed of sound due to the repulsive 3N interactions, which enter in $\Delta$-full EFT already at NLO. At N$^2$LO, we find again a similar behavior with the noticeable 3N contributions for all three interactions and the stiffest EOS given by the EMN N$^2$LO 500 interaction, while the EMN N$^2$LO 450 and $\Delta$NNLO$_{\textrm{GO}}\,(450)$ interactions yield more similar results. We also note that at N$^2$LO the results for $c_s^2$ are more similar for the three interactions and do not exhibit the stronger cutoff and interaction dependence at larger densities, as was observed for the Landau parameters and effective mass. This is because for $c_s^2 \sim (1+F_0)/(1+F_1/3)$ the $F_0$ and $F_1$ behaviors partly cancel. An important topic for future work will be to explore GP-Bayesian uncertainties for observables in Fermi liquid theory. Since both $m^*/m$ and $c_s^2$ are quantities obtained by (functional) derivatives from the total energy, the EFT truncation uncertainties will be larger. This is also evident from the order-by-order behavior in Fig.~\ref{fig:mEff_EFT_order}.

Finally, we study the MBPT convergence for $c_s^2$ in Fig.~\ref{fig:cs_MBPT_table}. Comparing the HF to MBPT(2) results (dashed vs. solid lines), we observe a well-converged behavior for $c_s^2$ for all interactions. The MBPT convergence for $c_s^2$ is better than for $m^*/m$, which follows the behavior of $F_1$ in Fig.\ref{fig:F1_MBPT_table}. We again attribute this to compensating effects from $F_0$ and $F_1$ for $c_s^2$. For a more complete uncertainty quantification, it will be important to develop MBPT order-by-order uncertainties and include these in combined uncertainty estimates.

\section{Summary and outlook}
\label{chap:summary}

We have presented results for the EOS of PNM and SNM up to densities $2 n_0$ based on different chiral NN+3N interactions, including new low-resolution interactions from \citet{Arthuis:2024mnl} and comparing $\Delta$-full to $\Delta$-less interactions at N$^2$LO. Our calculations are based on an expansion around HF up to third order in MBPT. For all interactions, our MBPT results of the EOS show a good many-body convergence for all densities studied, with the best convergence for the SRG-evolved interactions or for those with lower cutoffs, as expected. In addition to the MBPT uncertainties estimated from the MBPT(3) corrections, we assessed the EFT truncation uncertainties using the GP-Bayes methods from Refs.~\cite{Drischler2020PRL,Drischler2020}. While the $\Delta$-full interactions exhibited an improved order-by-order convergence due to 3N forces entering already at NLO, we found no improvement in the EFT uncertainties at N$^2$LO of $\Delta$-full versus $\Delta$-less interactions. In general, we found the EFT truncation uncertainties to dominate, but for a full uncertainty quantification, a framework that allows a joint analysis of chiral and many-body uncertainties is needed.

We have used the SNM and PNM results to extract ranges for the incompressibility $K$, the symmetry energy $E_\text{sym}$, and the $L$ parameter. This resulted in a broad range for the incompressibility $251 \mev \leqslant K \leqslant 352 \mev$, while the symmetry energy is tightly constrained to $32.0 \mev \leqslant E_{\text{sym}} \leqslant 33.9 \mev$. The $L$ parameter was lower for the low-resolution interactions $46.7 \mev \leqslant L \leqslant 51.5 \mev$ and could reach up to $L \leqslant 67.3 \mev$ for the other interactions considered. These ranges only provided an estimate of the Hamiltonian uncertainty, so that also here combining correlated EFT uncertainties and many-body uncertainties is important future work.

We have then extended our MC framework to calculate all NN and 3N contributions to the quasiparticle interaction up to second order in MBPT. In this work, we focused on PNM and on the spin-independent quasiparticle interaction and the Landau parameters $F_0$ and $F_1$, as these are most relevant to the EOS. In the future, we will extend our work to other Fermi liquid parameters, including noncentral contributions~\cite{Schwenk:2003bc}, as well as to SNM and asymmetric matter. We have used our EOS calculations to benchmark, at the same MBPT order, the speed of sound calculated from the Fermi liquid parameters $c_s^2 \sim (1+F_0)/(1+F_1/3)$ against the thermodynamic derivative of the EOS $c_s^2 = \partial P/\partial \mathcal{E}$.

We have found a very good MBPT convergence for the speed of sound, with the MBPT(2) results very similar to HF for all interactions and densities considered. This clearly showed the important 3N repulsion coming already from the HF level. At $2 n_0$ the speed of sound squared ranges from $0.04$ for the softest [1.8/2.0 (EM)] to $0.16$ for the hardest (EMN N$^2$LO 500) interaction. For the Landau parameters $F_0$ and $F_1$ individually, we found very similar results for the different interactions up to $0.1 \fmiq$ and a slower MBPT convergence for densities above $n_0$. This resulted in a broad possible range of the effective mass towards higher densities (as estimated from the Hamiltonian uncertainty), with all interactions leading to an increasing effective mass $m^*/m > 1$ above $0.2 \fmiq$. This effective mass behavior has very interesting effects on the thermal properties of the EOS~\cite{Keller2021,Huth:2020ozf,Keller2023} and impacts supernova and merger simulations~\cite{Yasin:2018ckc,Schneider:2019shi,Jacobi:2023olu}. The harder interactions explored in this work (EMN N$^2$LO 450/500) even yield effective masses as high as $m^*/m \approx 1.6-2.3$ at $2 n_0$. In order to assess the many-body convergence, going to MBPT(3) for the quasiparticle interaction would be desirable. This will, however, require automatized diagram generation techniques. Finally, extending the FLT calculations and consistent EOS calculations to N$^3$LO is work in progress.

\vspace*{5mm}

\section*{Acknowledgments}

We thank Hannah G\"ottling for the calculations of the EFT truncation uncertainties using GP-Bayes methods and Pierre Arthuis and Alexander Tichai for providing matrix elements. We moreover thank all three as well as C.J. Pethick for discussions. In addition, we thank Francesco Marino for providing benchmark results for our nuclear matter calculations and Andreas Ekström for providing matrix elements of the NNLO$_{\text{sim}}$ interactions. This work was supported in part by the European Research Council (ERC) under the European Union’s Horizon 2020 research and innovation programme (Grant Agreement No.~101020842) and by the Deutsche Forschungsgemeinschaft (DFG, German Research Foundation) -- Project-ID 279384907 -- SFB 1245.

\section*{Data availability}

The data that support the findings of this article are openly available ~\cite{zenodo:data}.

\bibliography{literature}

\end{document}